\documentclass[preprint,review ,authoryear,11pt]{elsarticle}

\usepackage{amssymb,multirow,amsmath,graphicx,array,adjustbox,booktabs,hyperref,rotating,soul,url}
\usepackage[no-weekday]{eukdate}
\usepackage{geometry}
\usepackage[usenames,dvipsnames]{color}

\usepackage[normalem]{ulem}

\usepackage{array}
\newcolumntype{L}[1]{>{\raggedright\let\newline\\\arraybackslash\hspace{0pt}}m{#1}}
\newcolumntype{C}[1]{>{\centering\let\newline\\\arraybackslash\hspace{0pt}}m{#1}}
\newcolumntype{R}[1]{>{\raggedleft\let\newline\\\arraybackslash\hspace{0pt}}m{#1}}

\usepackage{mathpazo,parskip}
\begin{document}

\begin{frontmatter}

\title{Forecasting and its Beneficiaries}

\author[label1]{Bahman Rostami-Tabar \corref{cor1}}
\ead{rostami-tabarb@cardiff.ac.uk}
 \cortext[cor1]{Correspondence: B. Rostami-Tabar, Cardiff Business School, Cardiff University, Cardiff, CF10 3EU, UK. Tel.: +44-(0)29 2087 0723}
 \address[label1]{Cardiff Business School, Cardiff University, Cardiff, CF10 3EU, UK}
 \author[label2]{John E. Boylan}
 \address[label2]{Department of Management Science, Lancaster University, Lancaster, LA1 4YX, UK}

\begin{abstract}
This chapter addresses the question of who benefits from forecasting, using 'Forecasting for Social Good' as a motivating framework. Barriers to broadening the base of beneficiaries are identified, and some parallels are drawn with similar concerns that were expressed in the Operational Research literature some years ago.

A recent initiative, called `Democratising Forecasting', is discussed, highlighting its achievements, challenges,  limitations and  future agenda. Communication issues between the major forecasting stakeholders are also examined, with pointers being given for more effective communications, in order to gain the greatest benefits.
\end{abstract}


\end{frontmatter}

\section{Background} \label{sec:intro}

Forecasting informs critical decisions 
in public, private and third-sector organisations. Forecasts are used widely in such diverse areas as finance and economics, public policy, healthcare, transportation, supply chains and engineering. They can enable organisations to manage risk and deal with uncertainty.

A few examples illustrate the  scope of forecasting. In public policy, forecasts can enhance the quality of decision making 
by predicting changes in economic and social factors and their effects on the costs and benefits of policy options. Forecasting plays an important role in guiding government policy as well as business decisions regarding environmental challenges such as climate change, in situations ranging from CO$_{2}$ emissions to natural disasters and early warning systems \citep{calder2018computational}. Forecasting plays a crucial role in finance and economics, by anticipating how economies may evolve and informing the management of risk accordingly. In business and manufacturing, better forecasting can enable a more efficient supply chain and greater productivity. Forecasting is routinely used in the retail sector to ensure the availability of required items on the shelves.

In the forecasting literature, the main emphasis has been on the methods and techniques of forecasting. There is much less discussion on the questions of what purposes forecasts should serve and who should benefit from forecasting. If there is any discussion, the emphasis is on profit and other economic measures. \cite{rostami2021forecasting} have challenged the notion that forecasting must necessarily serve economic or financial aims. They point to the potential of `Forecasting for Social Good' (FSG). They characterise FSG as forecasting that contributes towards strengthening social foundations and respecting the ecological ceiling. These social and ecological impacts can be assessed both locally and globally. Inspired by the FSG agenda, we  ask who benefits from improved forecasting, and we discuss means by which more people can benefit.

Debate about the beneficiaries of modelling is not new. This issue was discussed in the Operational Research literature in a series of papers published in the 1970s and 1980s (e.g. \cite{bevan1978contribution}, \cite{dando1981kuhnian}, \cite{rosenhead1982materialist}). 
Power was a common theme underpinning those papers. The authors noted that it tended to be more powerful organisations that benefit from OR, and OR projects had powerful people in these organisations as their clients.

If OR in mainly conducted on behalf of powerful people in powerful organisations, then it is natural that OR modelling will serve their interests. This is sometimes observed quite explicitly in a model with 'maximisation of profit' as its objective. Forecasting models, on the other hand, may appear to be more neutral. Surely, everyone would be supportive of 'minimising forecast error' (if appropriately measured) as an objective. This is to miss the fact that forecasting is generally part of a broader set of activities, whose objectives may not be shared so universally.

Many processes have indirect beneficiaries, and these beneficiaries may be internal or external to the organisation. Take an example of a forecasting system to predict the demand on an Accident and Emergency department of a hospital. The direct beneficiary of the system is the person (or people) with responsibility for using those forecasts to plan work rosters of doctors, nurses and allied health professionals. If this is well done, then the indirect beneficiaries are those health professionals for whom work is planned, and those patients for whom treatment is being planned. The importance of indirect beneficiaries such as these should be recognised when designing any forecasting process.   

In this chapter, we examine the issue of forecasting beneficiaries from a number of perspectives. Firstly, we address the agenda of 'Forecasting for Social Good' and the different perspectives that may be called for. Then we move on to the major barriers to spreading the benefits of better forecasting more widely.
We review communication issues between stakeholders in forecasting. These issues are important in any setting, and must be addressed if real benefits are to be achieved and shared appropriately. We conclude by  examining a recent initiative to improve forecasting capabilities in less well-developed economies. The initiative is  called `Democratising Forecasting' and it aims to extend the range of organisations and users who can benefit from developments in forecasting methods and software. 

\section{Forecasting for Social Good}

What is the point of forecasting?  If asked this question, most academics and practitioners would say ‘to inform better decisions’, or words to that effect. This immediately prompts two further questions: i) what do we mean by better? and ii) better for whom?  Most of the organisations in which a more professional and analytical approach to forecasting is employed have economic and financial goals, such as minimising costs or maximising profits. Alternative goals may also be served by improved forecasting. \cite{rostami2021forecasting} defined ‘Forecasting for Social Good (FSG)’ as: ‘\textit{… a forecasting process that aims to inform decisions that prioritise the thriving of humanity over the thriving of economies by enhancing the social foundation and ecological ceilings that impact the public as a whole on both global and local levels}’. They remarked that FSG is inclusive and encompasses all organisations, irrespective of their industry or whether they are governmental, commercial or voluntary organisations. FSG can be considered as a self-contained area of inquiry that can lead to increased appreciation of forecasting as a worthwhile tool for various beneficiaries and their communities. The benefits should extend beyond the financial, although the forecasting process may also result in economic growth. It can still be considered as FSG if the main focus is to improve social and environmental conditions.

It is instructive to compare the concerns underpinning this definition of FSG with those of authors such as Bevan and Bryer, Dando and Bennett, and Rosenhead and Thunhurst, back in the 1970s and 1980s. Common to both is the disquiet that the less powerful members of society are not benefiting sufficiently from the rapid developments in modelling. The later work puts greater emphasis on global issues, although this had not been absent from earlier work, as summarised in the review of OR in developing countries by \cite{white2011developing}.  The later work also focuses more on environmental concerns, reflecting the growing awareness of ecological issues in recent years.

Traditionally, the success of forecasting been assessed purely in terms of forecast accuracy. \cite{boylan2006implication} argued that such accuracy measures should be complemented by accuracy-implication metrics. These measures assess the impact of forecast errors, for example on inventory costs and stock availability. \citep{rostami2021forecasting} developed this idea further by prioritising social and environmental measures within the set of accuracy-implication metrics.

As well as considering an extended set of performance metrics, it may also be necessary to use more qualitative approaches to forecasting. Data in less well-resourced organisations may be patchy or non-existent. In such cases, quantitative forecasting methods are less helpful and processes relying on judgement will become more prominent. There is an interesting parallel, here, with the Community OR movement in the UK, which emphasised the use of softer OR methods in voluntary or community-based organisations. In situations like these, forecasting processes should be designed to be as transparent and interpretable as possible. This promotes trust in the modelling process, particularly if there is a wider group of beneficiaries than might normally be the case.

In summary, `Forecasting for Social Good' may call for different perspectives on what constitutes success, and on the means of achieving success. Introducing these new perspectives in practice is not always straightforward. So, it may be fruitful to ask: what are the barriers to sharing the benefits of forecasting?

\section{Barriers to Sharing the Benefits of Forecasting}

In this section of the chapter, we review the barriers to 'Forecasting for Social Good', show how some are reducing in potency and point to ways of addressing the other barriers.

We suggest that there are three significant barriers to extending the benefits of forecasting: i) lack of access to resources (e.g. software, people), ii) lack of access to expertise and iii) poor communication between the various stakeholders involved in forecasting (including the decision-makers). The first barrier is particularly acute for resource-poor organisations but we shall argue that it is declining in significance. The second barrier is very important for poorer organisations and communities, and will be a major theme of this chapter. It actually affects richer organisations, too, who have the resources to invest in developing forecasting expertise, but fail to do so adequately. The final barrier is relevant to all organisations. It is sometimes under-emphasised in richer companies, to the detriment of their forecasting processes. It understandable that concerns about resource issues may draw attention away from communication in poorer organisations, but to neglect this vital issue would be misguided.

\subsection{Access to resources}

In the forecasting process, a number of resources - from skilled forecasters to specialist software - are required. Without them, forecasting model design and implementation become difficult \citep{levee1992key}. To acquire these resources, funds are needed to purchase software, recruit appropriate staff, 
and to educate stakeholders to increase awareness of the benefits of better forecasting. Unfortunately, the cost of the forecasting software, and of gaining 
modelling and statistical expertise can become very large \citep{sanders2017forecasting}.

The development of free open-source forecasting software over the past decade has been a breakthrough in making forecasting accessible to wider beneficiaries. This is because it can be installed and used at no cost to the user, while having huge support from the community of users, maintainers and developers. The most widely used open-source forecasting software is the forecast package for R \citep{hyndman2020package}, first released in 2006, and downloaded over 2 million times in 2019. Several other R packages for forecasting are listed on the CRAN Task View for Time Series \citep{TimeSeriespkg}. Another high level programming language that has been used to create open-source forecasting software is Python. The Statsmodels library \citep{seabold2010statsmodels} in Python allows for statistical forecasting and the scikit-learn library \citep{garreta2013learning} is used more for machine learning. The introduction of free open-source forecasting software has also led to the publication of free online educational resources (e.g. \cite{hyndman2021forecasting}, \cite{ twmr2021}), which can also play a crucial role in learning 
about forecasting theory and practice.

Although the dissemination of open-source software has made sophisticated forecasting methods more accessible, there are other resourcing issues to be resolved. The most obvious is the cost of recruiting and retaining appropriate staff to work in a forecasting or planning capacity. Those who have power in organisations are in a position to make decisions relating to staffing levels and the deployment of staff. 
These decisions can have a significant impact on the success, or otherwise, of the  implementation of new  forecasting processes. 
Without the support of powerful people in an organisation and their commitment to making necessary resources available, the implementation of a forecasting process will fail and its potential benefit will be lost.

\subsection{Access to Expertise}

Although open-source developments are fundamental in making forecasting available to wider beneficiaries, 
they may also introduce some 
problems into practice. Anyone can easily find packages that let them produce forecasts with just a few lines of code. Many Machine Learning (ML) algorithms provide repositories of forecasting models, automate the hyperparameter tuning process, and sometimes offer a way to put these models into production. The availability of such packages has led many people to think that the forecasting process can be fully automated, eliminating the need for forecasting expertise or interpretation. 
In fact, using forecasting models in practice requires various technical and non-technical skills that are lacking in many organisations. This becomes even more challenging in public sector, third sector and almost any sector in resource-poor environments in low/middle income countries. These are communities that may have benefited less from forecasting, and they may end up making poor decisions if there is a lack of expertise in forecasting   \citep{rostami2021business}.

A rigorous forecasting process is not just about running code. The process includes discussing the context with stakeholders, understanding decision making, identifying forecasting requirements, setting up the right evaluation metrics for the problem, including domain knowledge, forecasting model testing, and gaining feedback from stakeholders, which all might be very specific to each forecasting problem \citep{hyndman2021forecasting}. Moreover, forecasts are not always produced using statistical or ML models relying on statistical software. In fact, in less developed countries (low/lower income) where the quality of the data may be an issue, relying on structured management judgement including the Delphi method, forecasting by analogy, surveys, scenario forecasting and other judgemental forecasting approaches can be more relevant \citep{ALTAY2020}.

Therefore, educating forecasters on the entire forecasting process, so that they are able to understand what lies behind the forecasting models, the application of forecasting in various domains, and their implementation in practice, is imperative. This can help them to make better forecasts that are truly effective. Someone who just runs code might not be able to add value to real world problems in organisations. We need to go beyond training and educate the potential users of forecasts on their impacts and how they can be used to inform decision making process. Without such education, the benefits of forecasting cannot be assured.

\subsection{Communication Issues}

Even if finance is available to put software resources in place, and forecasters and planners have been recruited, this is still not a guarantee of success. It is also crucial that the right software is used, which is most appropriate to the needs of the organisation. Clear communication between software vendors and decision makers, in both directions, is crucial here. Otherwise, and unless all needs are met by freeware, expensive mistakes can be made.

It is also very important that forecasts are being generated at the right frequency, over appropriate horizons, at the right level of aggregation (e.g. at product group or Stock Keeping Unit level) and are evaluated using appropriate criteria. Evaluations, of both forecasting accuracy and forecasting utility, need to be able to take into account the range of possible outcomes that may be expected. All of this  requires effective communication channels between forecasters, planners and decision makers.

In the sections that follow, we examine communication issues from different perspectives. The successful resolution of these issues is essential if knowledge and skills are to be developed and grown in organisations, thereby allowing them to benefit to the fullest extent from expertise in forecasting. It should be stated that effective communication channels are a necessary but not sufficient condition for success. Internal politics brings different kinds of instability within a forecasting process, especially where there are many human interventions. Each intervention becomes an opportunity  to introduce bias and unnecessary inaccuracies \citep{makridakis2020benefits}. Different stakeholders may have different political interests within the organisation that skew the forecasting process. Powerful people may seek confirmation from the forecasting outputs and if they do not get that confirmation, they tend to ignore it \citep{cipriano2014power}. These issues are not unique to forecasting; they can affect other types of OR modelling too. To address these issues requires careful consideration of reward systems within an organisation and the promotion of a collaborative approach between different sectional interests.

\section{More Effective Communications} 

It would be a mistake to think of organisations as homogeneous entities. There are often different people, with different roles, involved in forecasting, planning and decision making activities. This is true not only in developed economies, but in developing ones as well.  

Considered as a whole, a forecasting system can be viewed as a human activity system \citep{checkland1981systems}, embedded within a wider decision making and monitoring system. In describing such a human activity system, three types of people are relevant: i) actors – those that undertake the activity; ii) customers – those for whom the activity is undertaken, and iii) owners – those with the power to start or stop a system. 
The inclusion of the third, and most powerful, category of people recognises that human activity systems require decisions to be taken about investment in resources (e.g. hardware and software), processes to be followed (including data capture, reporting and monitoring) and deployment of staff (to specific forecasting and planning tasks). The responsibility for organising forecasting processes and tasks may be delegated by owners to less senior managers. Resource decisions, unless relatively trivial, are not usually delegated. For owners to take such decisions, they must have access to the necessary funds and to have some appreciation of the benefits that can accrue from better forecasting. Lack of funds or lack of this appreciation and understanding can inhibit progress in forecasting and planning.

Checkland describes the second category of people as beneficiaries/victims, in a wry recognition that those who are said to benefit from a system may not do so in practice. 
The first category of people, the actors, play an essential role in the successful implementation of any forecasting system. 
The need to develop the knowledge and skills of these people has not always been fully recognised. It is certainly an issue in less developed economies, and this theme will be picked up in the next section of this chapter. It can be an issue in more advanced economies too. Forecasting education is often neglected in developed countries such as the United Kingdom. Our experience of running training courses in the UK is that demand planners can quickly absorb the principles of the methods they have been applying `blind` and can use their forecasting systems much more effectively as a result.

For forecasting to be beneficial in organisations, effective communication between the `actors', customers' and 'owners' is essential. However, the scope should be even wider. In this section, communications between three groups of 'forecast suppliers' (loosely defined) and two groups of 'forecast recipients' will be considered. The forecast suppliers are defined broadly to embrace forecast producers, software developers and forecasting academics. (The term 'forecasting practitioners' is sometimes used to cover the first two categories). The forecast recipients include both the direct recipients (the 'customers') and indirect recipients (the 'owners').

\subsection{Forecast Producers} 

It is important to recognise two important factors to increase acceptance and use of forecasting in practice: i) how forecasts should be communicated with stakeholders and ii) what should be communicated. Traditionally, forecasts are communicated as point forecasts (single numbers), which does not provide the information required for a decision maker to manage risk. Communicating forecasts as forecast distributions will be more beneficial in that regard. Moreover, communicating not only the distributions of the forecast variable but also their impact on decisions is critical. Considering an Accident and Emergency department, providing the forecast distribution of the future patient admissions using a forecasting model is important. However it is not something that a decision maker can take action on. Complementing this information with the distributions of costs, number of staff required and waiting times resulting from the demand forecasts is more beneficial.

For communications to be effective, forecasters need to have a good awareness of how forecasts should be displayed visually. This is especially important when communicating forecast distributions, so that owners and customers (using Checkland's terminology) can appreciate the information that they are being given. Later, we shall discuss the development needs of forecast recipients. A little training should go a long way to improving communications of forecasts and plans.

\subsection{Commercial Software Developers}

Commercial software developers have a keen interest in communicating with system owners, as they have the power to commission and purchase a new forecasting system. This decision needs to take into account numerous factors, including interfaces with existing systems in general and Enterprise Resource Planning systems in particular.  

Communication often focuses on factors other than forecasting accuracy because these issues will often be uppermost in the minds of the system owners. Moreover, given the lack of knowledge about forecasting, discussed previously, it is usually not possible to have an in-depth discussion on this matter.   

When forecasting is discussed, system owners may be impressed by software developers' descriptions of the latest 'trendy' methods incorporated into their software, or by anecdotal evidence of forecasting accuracy improvements at other organisations.  The problem is that there is hardly any objective evidence on the comparative accuracy of different forecasting software packages. Some information on a range of software packages is available in the biennial forecasting software review conducted on behalf of the Institute for Operations Research and Management Science (INFORMS) and published in \textit{MSOR Today}. These surveys show comparisons of the features available in software packages but do not attempt to assess their forecasting performance.

After purchase, software developers often conduct training with forecasters and demand planners in the organisation. \cite{boylan2021intermittent} have pointed out that there is great variation as to what may constitute training. If the training does not go beyond tuition in clicking the right buttons, then there will be little understanding of what to do when the forecasts decline in accuracy. There is also an important distinction between using a package to produce an operational forecast and using it as a 'what-if' scenario planner. To make the best use of these planners, users need to have an appreciation of the alternative methods available in their software and the parameters that control them.

\subsection{Forecasting Academics}

Although forecasting academics are usually at least two steps removed from forecasting clients, their contributions can be genuinely beneficial if harnessed appropriately. Their main impact, to date, has been in developed economies. In the next section, an initiative is described linking academics to less developed economies. 

\cite{singh2016academia} pointed to a widening gap between academic forecasting and business forecasting and, regrettably, his observation is still relevant five years later. There are three gaps that underpin the problem, namely the knowledge, research and implementation gaps \citep{Boylan2016tango}.  

The knowledge gap arises because of the lack of understanding of even basic forecasting theory by many practitioners, and the lack of appreciation, by many academics, of the practical settings in which forecasting is conducted. Filling these gaps is essential for meaningful communication to be established. Some progress has been made in this direction by the establishment of outlets such as \textit{Foresight: the International Journal of Applied Forecasting}, which attracts pieces written in an accessible style by both practitioners and academics. Opportunities for face-to-face communications have been offered by the \textit{Foresight Practitioner Conferences} and by events run by the \textit{Institute of Business Forecasting and Planning}. Also, some universities offer training courses and events specifically targeted at practitioners. However, such publications and events attract the attention mainly of forecasting practitioners, and often do not reach forecasting clients or system owners. Although these groups do not need a detailed appreciation of forecasting techniques, they do need to understand how to set realistic forecasting accuracy targets and to appreciate how the forecasts should inform their decision making. New communication channels are required to address this need. 

The research gap arises from academics putting too much emphasis on solving stylised problems, which can be somewhat remote from real business situations. This can prompt doubts from practitioners about the relevance of academic research to their organisations. More researchers are now testing innovations on real-world data. Some researchers have gone further and evaluated the business implications of forecasting errors from different methods. For example, \cite{wang2016select} examined the inventory implications of different forecasting approaches, including judgemental ones (often utilised in practice) on real business data. More research projects of this type are needed, together with good quality review articles and books, which synthesise the available empirical evidence.

The implementation gap refers to the delay between new methods being published and subsequently being adopted in software. It is here that software developers play a crucial role. When methods are adopted in commercial software, then their use follows soon after. Communication between academics and software developers has been enhanced in recent years by more researchers developing open-source software, in languages such as R and Python, to complement their academic papers. This is being picked up by some of the more adventurous developers, who are incorporating these new methods into their own commercial software offerings.

\section{Democratising forecasting}

While there are many initiatives such as Data Science for Social Good \citep{ghani2018data}, Artificial Intelligence for Social Good \citep{shi2020artificial}, Pro Bono Operations Research \citep{midgley2018community}, Statistics for Social Good \citep{hwang2019improving} and Forecasting for Social Good \citep{rostami2021forecasting} that investigate the potential benefits of using data, modelling and forecasting to tackle societal and environmental challenges, very few focus on the educational issues.
\cite{mcclure1981educating} argued that the lack of educational processes for the potential beneficiaries of Operational Research (OR) and Management Science (MS) caused difficulties in securing benefits from OR/MS approaches in practice. In particular, the importance of education for the users of OR -  those who do not necessarily develop models but use them in the decision making process - was highlighted.

\subsection{Data Science Initiatives}

With the increase in data availability in the last decade and the interest in using the power of data,
new data-driven areas such as Data Science (DS) have emerged \citep{cuquet2017societal}. \cite{hill2017democratizing} argued that it is important to make access to DS tools more widespread to ensure that end users benefit from them. To that end, they have designed a series of workshops and courses with the goal of giving users in online communities the ability to ask and answer their own questions and to build their skills to engage with other analysts and analyses. They also aim at reducing inequality by training under-represented groups in the field such as women, minorities, people with disabilities, and veterans. Workshops are offered at times, and at a cost, that makes participation by diverse groups of people possible.

The National Institutes of Health (NIH) has launched the Big Data to Knowledge (BD2K) initiative to facilitate both in-person and online learning, and open up the concepts of Data Science to the widest possible audience in the biomedical sciences. BD2K has created the Educational Resource Discovery Index (ERuDIte), which identifies, collects, describes, and organises online data science materials from various online sources to democratise novel insights and discoveries brought forth via large-scale data science training \citep{van2018democratizing}. These initiatives cover a wide range of topics and may include forecasting but not necessarily so. 
There are some other initiatives that democratise access to Data Science such as R-Ladies \citep{bellini2020r}. The focus here is to make Data Science, using R programming, accessible to people of genders currently underrepresented in the community worldwide. The organisation is articulated into 'chapters', groups hosting events in cities or remotely, the latter for the benefit of everyone, regardless of geographic location or personal circumstances. The focus of the group is to democratise programming skills in R via meetups (in-person and online), the abstract reviewers' network, the Slack channels and the mentorship programme. These initiatives are meeting with some success. Their main emphasis has been on broadening access within developed countries. Their principal focus has not been on developing countries.

\subsection{Democratising Forecasting Initiative}

Given the importance of educating stakeholders in order to increase the benefits of forecasting to wider communities, one of the co-authors of this book chapter launched the Democratising Forecasting (DF) initiative  \citep{democratisingforecasting2020} in 2018, supported by the  \textit{International Institute of Forecasters (IIF)}. The goal of this project is to provide forecasting education to individuals in developing countries around the world. This is born from a recognition of the benefits that forecasting knowledge can bring to advancing people's use of forecasting to inform decisions. However, it goes one step further by emphasising direct capacity building in deprived economies by educating future forecasters. This project aims to build skills, increase the systematic use of forecasting,  and engage with stakeholders to make better decisions. 

This work has involved designing a  curriculum, and then running a series of workshops delivered to academics (students and lecturers) and practitioners in developing countries. The three-day training course concentrates on the foundations of forecasting, using R, for forecasters, analysts and modellers, focusing on the theoretical background of forecasting methods, their benefits and limitations and their implementations and use in R software \citep{hyndman2020package}. The main emphasis of the workshops has been on educating learners about generating forecasts. Therefore, the primary audiences of the workshop have been academics and forecasters.

These workshops are used to explore the potential of, and challenges around, democratising forecasting. In designing and delivering the workshop, the intention is to broaden participation along several dimensions including geographical location, demographic characteristics, and academic fields. Partners from academia, research institutes or industry are selected to coordinate and help organising the workshops. Partners are selected based on World Bank data from low, low-middle and middle income countries.

To make the workshops more widely accessible, the training comes at no cost for participants or the hosting organisation, thanks to sponsorship from the International Institute of Forecasters. This is important for participants from lower income countries, who might not be able to afford workshop fees. Beginners are targeted and, accordingly, participants are not assumed to have any prior forecasting knowledge or any programming experience in R. The training is designed around a project based on producing forecasts for hospital admissions, to inform decisions regarding resource allocation. Participants spend the majority of their time in the sessions identifying decisions that require forecasts, determining what to forecast and the associated data requirements, analysing data, writing code for producing forecasts and evaluating accuracy. 
In order to enrich the participants' learning experience, various guest speakers are invited from international organisations such as the National Health Service (NHS), the International Committee of the Red Cross (ICRC), the Australian Bureau of Agricultural and Resource Economics and Sciences and SAP to talk about specific forecasting topics. The topics range from retail forecasting to agricultural and food forecasting, forecasting for emergency services and forecasting for humanitarian operations.
These talks provide a unique opportunity for learners in developing countries to engage with forecasters from prominent organisations around the world via an online platform. This creates a connection to the use of forecasting in the real world, as speakers share best practices on how forecasts are created and used and discuss challenges and practices that should be avoided.     

\subsection{Benefits and Impact of Democratising Forecasting}

The aim of the Democratising Forecasting project is to reach wider communities with limited access to high quality education, 
help participants to develop skills, increase awareness of the potential benefits of forecasting and 
empower participants to facilitate workshops in their country and create change.

The initial goal was to train 400 individuals, over 5 years in 20 countries. Nine workshops had been delivered in seven countries in Africa, the Middle East and South Asia by the end of 2019. Overall, 210 people have been trained, including undergraduate students, postgraduate students, academics and practitioners.
An evaluation feedback survey has been sent to learners at the end of each workshop to assess whether the initiative has achieved its stated aims and had an impact so far. While it might not be easy to measure the success of the  initiative using a post-workshop survey, we can provide some evidence from participants' feedback, which may serve as indications of the potential impact.

During the workshop, participants learn forecasting and R programming skills. They learn the theory and concepts of forecasting and how to implement methods in R. A participant explained how the workshop helped him to build skills in forecasting and R programming:
"\textit{The workshop helped me to manipulate and prepare data for forecasting and use R to generate forecast and evaluate its quality. I discovered how powerful R is for forecasting modeling.}"

The workshop also emphasises the use of forecasting for planning and decision making through various examples and guest speakers from different sectors. Although not relating directly to their own experience, this provides an opportunity for learners to think critically  about the role of  forecasting and its potential benefits for their organisations and wider society.
A participant highlighted how the workshop increased her awareness about the potential use of forecasts:
"\textit{I couldn't have imagined a link between forecasting and social good utilities. You have shown that link and I thank you for that.}"

Democratising forecasting workshops are delivered in person. They are coordinated with institutions in developing countries, who host and organise the workshop. Tutors from the UK travel to the country and deliver the training. By doing so, the initiative achieves its aim of  outreach to resource-poor countries where access to quality education might be an issue. \cite{rostami2021business} discussed the knowledge and expertise gap in the area of forecasting in developing countries and emphasised  that very little has been done over the past two decades to reduce this gap. A participant highlighted the importance of organising the forecasting workshop in her country:
"\textit{Not everyone in the world has the privilege or the chance to get a proper education and it is us, those who had this privilege, [who give] thanks for making the forecasting workshop accessible to us}."

Ideally, the democratising forecasting initiative should empower participants and ultimately bring changes not only at the personal level but more widely in terms of enabling decisions to be more well informed. 
While it is difficult to measure the impact of the workshop on the latter, there are indications of the former from end-of-workshop  feedback. A participant reported that the workshop had led to a significant change in her priorities for her professional life:

"\textit{Honestly, this workshop made me reset some of my priorities. In fact, I thought that I was determined and that I already chose the next steps in my career. Then, after this training, I found myself really interested in forecasting and especially forecasting for social good. So, I really welcome any opportunity to create the community that you talked about today and I would be honored to be part of it.}"

\subsection{Challenges in Delivering Democratising Forecasting Workshops}

Democratising Forecasting focuses on educating people about forecasting in developing countries and requires coordination with partners in those countries to make the workshops accessible to wider beneficiaries. Partners are  responsible for in-country organisation. Tasks include advertising the workshop among potential participants, registration, communication and providing a venue.

While the idea of Democratising Forecasting is based on the distribution of forecasting knowledge, at no cost, in developing countries, it would be an oversimplification to assume that everyone is open to the idea. In fact, one of the most important challenges in the Democratising Forecasting workshops has been finding partners who are genuinely interested in knowledge distribution in this area and who are able and willing to coordinate effectively to meet the learning requirements of the workshop. The coordination with partners is needed prior to the workshop delivery to ensure that learning will take place, and requires having the right learners in the workshop with the right level of infrastructure and support. In the following, some challenges related to coordination and delivery are highlighted.

It had been agreed with partners that the workshop should be accessible to any learner at no cost. When delivering the workshops, it became evident that some institutions made it accessible only to their own students and staff. In one case, the hosting organisation was asking for money to organise the workshop, to provide a room and internet access. The amount they proposed was clearly aimed at making a profit from organising the event. Some partners brought in irrelevant people (e.g. secretaries) to show off the high number of attendees. These examples 
revealed, very clearly, that the concept of sharing knowledge and making it accessible to wider communities at no cost is not always understood or appreciated in some institutions.

Another challenge that was observed is related to the physical environment. Research highlights the importance of the physical environment on learning \citep{brooks2011space}. This was neglected by partners in some countries, especially in Africa. In one case, a workshop had to be delivered in a room without any windows or ventilation. Also, the layout of the room did not allow for any group discussion and it was difficult for the tutor to move around and help students who may have been having issues in using the software.

Overall, it is evident that the lack of dedicated partners who share the value of the  Democratising Forecasting project can be an important barrier in achieving the project's aim.

\subsection{Limitations of the Democratising Forecasting Initiative}

One of the limitations of the Democratising Forecasting project has been the lack of assistance to learners after attending the workshop, to overcome the issues they face. 
This would include helping them  at the transitional stage from the workshop to real-world settings or to become a trainer. This is highly desirable but additional financial and people resources would be required to support learners in this way.

The practical aspects of the workshop could be delivered based on problems/datasets that participants are familiar with. While this might be helpful for the learning process, it would require skilled mentors to help the instructor and such mentors have not been available.

So far, the Democratising Forecasting (DF) project has been mainly focusing on the theory of forecasting models and educating analysts to produce forecasts using R, rather than on the role of forecasts in the decision making-process. This limitation may not be restricted to the DF project and could be a more general tendency in designing forecasting education programs, which may create issues in accepting forecasting as a useful tool. Similar issues have also been highlighted in the Management Science and Operations Research education literature \citep{mcclure1981educating,zahedi1985ms}.
Also, the workshop content only covers forecasting using statistical methods, which assumes data availability. However, there are many situations where there is limited capacity to record data in developing countries, which means the data might not be available or the collected data might not be reliable. 
Including sessions on judgmental forecasting can enhance the curriculum in that regard.

In such programmes, there is often insufficient discussion on forecasting beneficiaries, the link between forecasting and decision making and how forecasts should be used. This raises the important question of whether learning theories about forecasting and applying models on datasets is enough to prepare learners for forecasting in practice. In the examples discussed in the DF training courses, there is usually no role for the user of forecasts to play. This may create an issue in using forecasts in practice, as learners may conclude that this is the end and generated forecasts are the solution to the problem. Putting more emphasis on the relationship between forecasts and decision making in forecasting education programmes and educating the main beneficiaries of forecasting, such as managers and decision makers, is fundamental in encouraging the  systematic use of forecasting. It is intended that this aspect of Democratising Forecasting will become a focal point for future developments.

\subsection{Future Agenda for Democratising Forecasting}

Given the importance of judgemental forecasting in practice, especially in resource-poor environments where data is not available or is incomplete, and the fact that DF workshops have been covering only statistical methods, it is planned to introduce a new session on judgemental forecasting in future workshops. The sessions will introduce different approaches in judgemental forecasting, and there will be a case study focusing on scenario forecasting, given its relevance in many situations for contingency planning.

To overcome the lack of support and communication with DF participants after  workshops have finished, it is planned to create the Democratising Forecasting Slack channel and to invite attendees to join the channel. It will be used as a communication channel 
and, in particular, members can use it to post questions and/or answers. Previous attendees can also support new ones in their journey, help them to solve issues and create an online learning environment. The Slack platform has proved to be a very effective tool to assist learners, especially in data science related programmes \citep{perkel2017scientists, vela2018using}. It can be used as a single communication channel with attendees before, during and after attending a workshop. The aim is to use this channel to increase engagement, enhance the learning experience, and assist learners during their transition stage, after attending the workshop.

Aligning with the principles of Forecasting for Social Good, the future agenda is to encourage workshop participants to put what they have learned into practice, and to empower them to create change by using forecasting to make better informed decisions and policies. To this end, the Forecasting for Social Good Research Grant has been introduced \citep{fsgrg2021}. This grant will be awarded annually to researchers in developing countries as the project lead and its main objective is to improve the use of forecasting tools to inform decisions that prioritise the wellbeing of people and the planet.

Potential users of forecasts (e.g. managers, planners, decision makers) may not fully benefit from the current workshop, as it mainly focuses on forecasting methods and their implementation in R software. In recognition of this issue, it is planned to design and deliver a one-day workshop on forecasting for managers. 
The focus of this workshop will be on concepts and considerations about forecasting from the perspective of the consumers of forecasting, rather than the producers. The workshop will be aimed at increasing awareness of the potential benefits of forecasting for managers and decision makers and encouraging the systematic use of forecasts. The workshop will be coordinated and hosted in developing countries by 
a leading international humanitarian organisation.

\section{Conclusions}\label{sec:concl}

In many modern organisations, multiple decisions are 
made in the light of predictions generated by forecasting processes. A forecasting process should be perceived as successful if it enables better decisions to be made, bringing benefits to organisations and wider society. This requires a shift from the way forecasting success is assessed, which has been traditionally based on forecast accuracy alone. The potential benefits resulting from the use of forecasting should include not only economic outcomes but also social and environmental measures at both local and global levels. This would lead to an increased appreciation of forecasting as an activity with a much broader base of beneficiaries than has previously been contemplated.

Ensuring that forecasting benefits decision making requires far more than developing the latest forecasting models, which has been the main emphasis in the academic forecasting literature. There are various factors that may act as barriers to realising the benefits that can arise from forecasting. We have discussed some of the principal reasons why benefits might not be shared, including lack of access to resources and expertise, and poor communication between the various stakeholders involved in the forecasting process. In addition to improving internal communications, it was argued that better communications are also needed with software developers and forecasting academics.

In recognition of the benefits that forecasting knowledge and skills can bring to organisations, and to overcome some of the barriers to realising these benefits, the Democratising Forecasting (DF) initiative has been established, to extend the range of users and organisations. This project builds capacity in developing countries by educating the potential beneficiaries of forecasting. Overall, the project has been a success so far. It has reached out to communities in lower/lower-middle/middle income countries, helping learners to  develop forecasting and software skills, empowering them to create change and increasing awareness of the potential benefits of forecasting.

Reflecting on the experience of delivering workshops in the Democratising Forecasting project, we highlighted some issues and challenges that may have some broader relevance. It was concluded that perhaps the most important factor in making an international learning project a success is having a dedicated partner to coordinate and organise the workshop in the country. Without such a partner, all efforts to  design a high-quality programme and to make it more widely accessible, may prove fruitless.

In DF workshops, the main emphasis has been on  statistical methods and/or software. However, there are many situations where data might not be available or incomplete, and this is where judgemental forecasting becomes essential. This will be included in future workshops and should, indeed, be a part of any business forecasting educational programme. We also recognise that it is critical to assist learners after attending a workshop in their transition to produce and use forecasts in the real world. To that end, a grant has been established, which helps participants to put in practice their learning in the forecasting for social good area. 

Further reflection brought the realisation that the focus of the DF workshops has been mainly from the perspective of forecasters or forecasting modellers (i.e. on producing forecasts and using the software) rather then the perspective of the decision makers. This omission is important, given their crucial role in adopting, gaining acceptance and sustaining new forecasting processes. New forms of engagement with these decision makers are currently being planned. More generally, forecasting education should recognise the important role of decision makers as the main beneficiaries and the relationship between forecasting and decision making should be covered in any programme.

In similar initiatives such as Data Science for Social Good, there are indications of success such as peer review publications, number of organisations involved, number of project completed, and  number of fellowship awarded. Democratising Forecasting has also been growing in term of outreach and number of participants benefiting from its workshops, since their launch in January 2018. However, a full set of measures has not yet been developed and so it is difficult to assess its impact objectively. 
Success might be best evaluated based on actions taken, and their effects, as a result of what was learned during the workshop and afterwards. 

We believe that there is a broad demand for forecasting skills in communities that may have benefited less from forecasting, such as the public sector, third sector and almost any sector in resource-poor environments in low/middle income countries. We hope that we have  provided a vision of what a democratised forecasting curriculum might look like. We believe that coupling Democratising Forecasting with Forecasting for Social Good creates opportunities for societal and environment benefits beyond what was been achieved so far. 

\clearpage

 \bibliographystyle{elsarticle-harv}
\bibliography{referencesrostamitabar}

\begin{thebibliography}{39}
\expandafter\ifx\csname natexlab\endcsname\relax\def\natexlab#1{#1}\fi
\providecommand{\url}[1]{\texttt{#1}}
\providecommand{\href}[2]{#2}
\providecommand{\path}[1]{#1}
\providecommand{\DOIprefix}{doi:}
\providecommand{\ArXivprefix}{arXiv:}
\providecommand{\URLprefix}{URL: }
\providecommand{\Pubmedprefix}{pmid:}
\providecommand{\doi}[1]{\href{http://dx.doi.org/#1}{\path{#1}}}
\providecommand{\Pubmed}[1]{\href{pmid:#1}{\path{#1}}}
\providecommand{\bibinfo}[2]{#2}
\ifx\xfnm\relax \def\xfnm[#1]{\unskip,\space#1}\fi
\bibitem[{Altay and Narayanan(2020)}]{ALTAY2020}
\bibinfo{author}{Altay, N.}, \bibinfo{author}{Narayanan, A.},
  \bibinfo{year}{2020}.
\newblock \bibinfo{title}{Forecasting in humanitarian operations: Literature
  review and research needs}.
\newblock \bibinfo{journal}{International Journal of Forecasting} \URLprefix
  \url{https://www.sciencedirect.com/science/article/pii/S0169207020301151},
  \DOIprefix\doi{https://doi.org/10.1016/j.ijforecast.2020.08.001}.
\bibitem[{Bellini~Saibene et~al.(2020)Bellini~Saibene, Vitolo, LeDell, Frick
  and Acion}]{bellini2020r}
\bibinfo{author}{Bellini~Saibene, Y.}, \bibinfo{author}{Vitolo, C.},
  \bibinfo{author}{LeDell, E.}, \bibinfo{author}{Frick, H.},
  \bibinfo{author}{Acion, L.}, \bibinfo{year}{2020}.
\newblock \bibinfo{title}{R-{L}adies {G}lobal, a worldwide organisation to
  promote gender diversity in the {R} community.}, in: \bibinfo{booktitle}{EGU
  General Assembly Conference Abstracts}, p. \bibinfo{pages}{20530}.
\bibitem[{Bevan and Bryer(1978)}]{bevan1978contribution}
\bibinfo{author}{Bevan, R.G.}, \bibinfo{author}{Bryer, R.A.},
  \bibinfo{year}{1978}.
\newblock \bibinfo{title}{On measuring the contribution of {OR}}.
\newblock \bibinfo{journal}{Journal of the Operational Research Society}
  \bibinfo{volume}{29}, \bibinfo{pages}{409--418}.
\bibitem[{Boylan and Syntetos(2006)}]{boylan2006implication}
\bibinfo{author}{Boylan, J.E.}, \bibinfo{author}{Syntetos, A.A.},
  \bibinfo{year}{2006}.
\newblock \bibinfo{title}{Accuracy and accuracy-implication metrics for
  intermittent demand}.
\newblock \bibinfo{journal}{Foresight: the International Journal of Applied
  Forecasting} , \bibinfo{pages}{39--42}.
\bibitem[{Boylan and Syntetos(2016)}]{Boylan2016tango}
\bibinfo{author}{Boylan, J.E.}, \bibinfo{author}{Syntetos, A.A.},
  \bibinfo{year}{2016}.
\newblock \bibinfo{title}{Commentary: It takes two to tango}.
\newblock \bibinfo{journal}{Foresight: The International Journal of Applied
  Forecasting} , \bibinfo{pages}{26--29}.
\bibitem[{Boylan and Syntetos(2021)}]{boylan2021intermittent}
\bibinfo{author}{Boylan, J.E.}, \bibinfo{author}{Syntetos, A.A.},
  \bibinfo{year}{2021}.
\newblock \bibinfo{title}{Intermittent Demand Forecasting: Context, Methods and
  Applications}.
\newblock \bibinfo{publisher}{John Wiley \& Sons}.
\bibitem[{Brooks(2011)}]{brooks2011space}
\bibinfo{author}{Brooks, D.C.}, \bibinfo{year}{2011}.
\newblock \bibinfo{title}{Space matters: The impact of formal learning
  environments on student learning}.
\newblock \bibinfo{journal}{British Journal of Educational Technology}
  \bibinfo{volume}{42}, \bibinfo{pages}{719--726}.
\bibitem[{Calder et~al.(2018)Calder, Craig, Culley, de~Cani, Donnelly, Douglas,
  Edmonds, Gascoigne, Gilbert, Hargrove et~al.}]{calder2018computational}
\bibinfo{author}{Calder, M.}, \bibinfo{author}{Craig, C.},
  \bibinfo{author}{Culley, D.}, \bibinfo{author}{de~Cani, R.},
  \bibinfo{author}{Donnelly, C.A.}, \bibinfo{author}{Douglas, R.},
  \bibinfo{author}{Edmonds, B.}, \bibinfo{author}{Gascoigne, J.},
  \bibinfo{author}{Gilbert, N.}, \bibinfo{author}{Hargrove, C.}, et~al.,
  \bibinfo{year}{2018}.
\newblock \bibinfo{title}{Computational modelling for decision-making: where,
  why, what, who and how}.
\newblock \bibinfo{journal}{Royal Society Open Science} \bibinfo{volume}{5},
  \bibinfo{pages}{172096}.
\bibitem[{Checkland(1981)}]{checkland1981systems}
\bibinfo{author}{Checkland, P.}, \bibinfo{year}{1981}.
\newblock \bibinfo{title}{Systems Thinking, Systems Practice}.
\newblock \bibinfo{publisher}{Wiley}, \bibinfo{address}{Chichester, UK}.
\bibitem[{Cipriano and Gruca(2014)}]{cipriano2014power}
\bibinfo{author}{Cipriano, M.}, \bibinfo{author}{Gruca, T.S.},
  \bibinfo{year}{2014}.
\newblock \bibinfo{title}{The power of priors: How confirmation bias impacts
  market prices}.
\newblock \bibinfo{journal}{The Journal of Prediction Markets}
  \bibinfo{volume}{8}, \bibinfo{pages}{34--56}.
\bibitem[{Cuquet et~al.(2017)Cuquet, Vega-Gorgojo, Lammerant, Finn
  et~al.}]{cuquet2017societal}
\bibinfo{author}{Cuquet, M.}, \bibinfo{author}{Vega-Gorgojo, G.},
  \bibinfo{author}{Lammerant, H.}, \bibinfo{author}{Finn, R.}, et~al.,
  \bibinfo{year}{2017}.
\newblock \bibinfo{title}{Societal impacts of big data: challenges and
  opportunities in {Europe}}.
\newblock \bibinfo{journal}{arXiv preprint arXiv:1704.03361} .
\bibitem[{Dando and Bennett(1981)}]{dando1981kuhnian}
\bibinfo{author}{Dando, M.R.}, \bibinfo{author}{Bennett, P.G.},
  \bibinfo{year}{1981}.
\newblock \bibinfo{title}{A {K}uhnian crisis in {M}anagement {S}cience?}
\newblock \bibinfo{journal}{Journal of the Operational Research Society}
  \bibinfo{volume}{32}, \bibinfo{pages}{91--103}.
\bibitem[{Garreta and Moncecchi(2013)}]{garreta2013learning}
\bibinfo{author}{Garreta, R.}, \bibinfo{author}{Moncecchi, G.},
  \bibinfo{year}{2013}.
\newblock \bibinfo{title}{Learning scikit-learn: machine learning in {P}ython}.
\newblock \bibinfo{publisher}{Packt Publishing Ltd}.
\bibitem[{Ghani(2018)}]{ghani2018data}
\bibinfo{author}{Ghani, R.}, \bibinfo{year}{2018}.
\newblock \bibinfo{title}{Data science for social good and public policy:
  examples, opportunities, and challenges}, in: \bibinfo{booktitle}{The 41st
  International ACM SIGIR Conference on Research \& Development in Information
  Retrieval}, pp. \bibinfo{pages}{3--3}.
\bibitem[{Hill et~al.(2017)Hill, Dailey, Guy, Lewis, Matsuzaki and
  Morgan}]{hill2017democratizing}
\bibinfo{author}{Hill, B.M.}, \bibinfo{author}{Dailey, D.},
  \bibinfo{author}{Guy, R.T.}, \bibinfo{author}{Lewis, B.},
  \bibinfo{author}{Matsuzaki, M.}, \bibinfo{author}{Morgan, J.T.},
  \bibinfo{year}{2017}.
\newblock \bibinfo{title}{Democratizing data science: The community data
  science workshops and classes}, in: \bibinfo{booktitle}{Big Data Factories}.
  \bibinfo{publisher}{Springer}, pp. \bibinfo{pages}{115--135}.
\bibitem[{Hwang et~al.(2019)Hwang, Orenstein, Cohen, Pfeiffer and
  Mackey}]{hwang2019improving}
\bibinfo{author}{Hwang, J.}, \bibinfo{author}{Orenstein, P.},
  \bibinfo{author}{Cohen, J.}, \bibinfo{author}{Pfeiffer, K.},
  \bibinfo{author}{Mackey, L.}, \bibinfo{year}{2019}.
\newblock \bibinfo{title}{Improving subseasonal forecasting in the western {US}
  with machine learning}, in: \bibinfo{booktitle}{Proceedings of the 25th ACM
  SIGKDD International Conference on Knowledge Discovery \& Data Mining}, pp.
  \bibinfo{pages}{2325--2335}.
\bibitem[{Hyndman(2020)}]{TimeSeriespkg}
\bibinfo{author}{Hyndman, R.J.}, \bibinfo{year}{2020}.
\newblock \bibinfo{title}{Time series {CRAN} task view}.
\newblock \URLprefix
  \url{https://cran.r-project.org/web/views/TimeSeries.html}.
\bibitem[{Hyndman and Athanasopoulos(2021)}]{hyndman2021forecasting}
\bibinfo{author}{Hyndman, R.J.}, \bibinfo{author}{Athanasopoulos, G.},
  \bibinfo{year}{2021}.
\newblock \bibinfo{title}{Forecasting: principles and practice}.
\newblock \bibinfo{publisher}{OTexts}, \bibinfo{address}{Melbourne, Australia}.
\newblock \URLprefix \url{http://OTexts.com/fpp3/}.
\bibitem[{Hyndman et~al.(2020)Hyndman, Athanasopoulos, Bergmeir, Caceres,
  Chhay, O'Hara-Wild, Petropoulos, Razbash, Wang and
  Yasmeen}]{hyndman2020package}
\bibinfo{author}{Hyndman, R.J.}, \bibinfo{author}{Athanasopoulos, G.},
  \bibinfo{author}{Bergmeir, C.}, \bibinfo{author}{Caceres, G.},
  \bibinfo{author}{Chhay, L.}, \bibinfo{author}{O'Hara-Wild, M.},
  \bibinfo{author}{Petropoulos, F.}, \bibinfo{author}{Razbash, S.},
  \bibinfo{author}{Wang, E.}, \bibinfo{author}{Yasmeen, F.},
  \bibinfo{year}{2020}.
\newblock \bibinfo{title}{{forecast}: Forecasting functions for time series and
  linear models}.
\newblock \URLprefix \url{https://pkg.robjhyndman.com/forecast/}.
  \bibinfo{note}{r package version 8.12}.
\bibitem[{Kuhn and Silge(2021)}]{twmr2021}
\bibinfo{author}{Kuhn, M.}, \bibinfo{author}{Silge, J.}, \bibinfo{year}{2021}.
\newblock \bibinfo{title}{Tidy Modeling with R}.
\newblock \URLprefix \url{https://www.tmwr.org/}.
\bibitem[{LeVee(1992)}]{levee1992key}
\bibinfo{author}{LeVee, G.S.}, \bibinfo{year}{1992}.
\newblock \bibinfo{title}{The key to understanding the forecasting process}.
\newblock \bibinfo{journal}{The Journal of Business Forecasting}
  \bibinfo{volume}{11}, \bibinfo{pages}{12}.
\bibitem[{Makridakis et~al.(2020)Makridakis, Bonnell, Clarke, Fildes,
  Gilliland, Hoover, Tashman et~al.}]{makridakis2020benefits}
\bibinfo{author}{Makridakis, S.}, \bibinfo{author}{Bonnell, E.},
  \bibinfo{author}{Clarke, S.}, \bibinfo{author}{Fildes, R.},
  \bibinfo{author}{Gilliland, M.}, \bibinfo{author}{Hoover, J.},
  \bibinfo{author}{Tashman, L.}, et~al., \bibinfo{year}{2020}.
\newblock \bibinfo{title}{The benefits of systematic forecasting for
  organizations: The {UFO} project}.
\newblock \bibinfo{journal}{Foresight: The International Journal of Applied
  Forecasting} , \bibinfo{pages}{45--56}.
\bibitem[{McClure(1981)}]{mcclure1981educating}
\bibinfo{author}{McClure, R.H.}, \bibinfo{year}{1981}.
\newblock \bibinfo{title}{Educating the future users of {OR}}.
\newblock \bibinfo{journal}{Interfaces} \bibinfo{volume}{11},
  \bibinfo{pages}{108--112}.
\bibitem[{Midgley et~al.(2018)Midgley, Johnson and
  Chichirau}]{midgley2018community}
\bibinfo{author}{Midgley, G.}, \bibinfo{author}{Johnson, M.P.},
  \bibinfo{author}{Chichirau, G.}, \bibinfo{year}{2018}.
\newblock \bibinfo{title}{What is community {O}perational {R}esearch?}
\newblock \bibinfo{journal}{European Journal of Operational Research}
  \bibinfo{volume}{268}, \bibinfo{pages}{771--783}.
\bibitem[{Perkel(2017)}]{perkel2017scientists}
\bibinfo{author}{Perkel, J.M.}, \bibinfo{year}{2017}.
\newblock \bibinfo{title}{How scientists use {S}lack}.
\newblock \bibinfo{journal}{Nature News} \bibinfo{volume}{541},
  \bibinfo{pages}{123}.
\bibitem[{Rosenhead and Thunhurst(1982)}]{rosenhead1982materialist}
\bibinfo{author}{Rosenhead, J.}, \bibinfo{author}{Thunhurst, C.},
  \bibinfo{year}{1982}.
\newblock \bibinfo{title}{A materialist analysis of {O}perational {R}esearch}.
\newblock \bibinfo{journal}{Journal of the Operational Research Society}
  \bibinfo{volume}{33}, \bibinfo{pages}{111--122}.
\bibitem[{Rostami-Tabar(2021a)}]{fsgrg2021}
\bibinfo{author}{Rostami-Tabar}, \bibinfo{year}{2021}a.
\newblock \bibinfo{title}{Forecasting for social good research grant}.
\newblock \URLprefix
  \url{https://forecasters.org/programs/research-awards/forecasting-for-social-good-research-grant/}.
\bibitem[{Rostami-Tabar(2020)}]{democratisingforecasting2020}
\bibinfo{author}{Rostami-Tabar, B.}, \bibinfo{year}{2020}.
\newblock \bibinfo{title}{Democratising forecasting}.
\newblock \URLprefix \url{https://forecasters.org/events/iif-workshops/}.
  \bibinfo{note}{sponsored by the International Institute of Forecasters}.
\bibitem[{Rostami-Tabar(2021b)}]{rostami2021business}
\bibinfo{author}{Rostami-Tabar, B.}, \bibinfo{year}{2021}b.
\newblock \bibinfo{title}{Business forecasting in developing countries}.
\newblock \bibinfo{journal}{Business Forecasting: The Emerging Role of
  Artificial Intelligence and Machine Learning} , \bibinfo{pages}{382}.
\bibitem[{Rostami-Tabar et~al.(2021)Rostami-Tabar, Ali, Hong, Hyndman, Porter
  and Syntetos}]{rostami2021forecasting}
\bibinfo{author}{Rostami-Tabar, B.}, \bibinfo{author}{Ali, M.M.},
  \bibinfo{author}{Hong, T.}, \bibinfo{author}{Hyndman, R.J.},
  \bibinfo{author}{Porter, M.D.}, \bibinfo{author}{Syntetos, A.},
  \bibinfo{year}{2021}.
\newblock \bibinfo{title}{Forecasting for social good}.
\newblock \bibinfo{journal}{International Journal of Forecasting} \URLprefix
  \url{https://www.sciencedirect.com/science/article/pii/S0169207021000510},
  \DOIprefix\doi{https://doi.org/10.1016/j.ijforecast.2021.02.010}.
\bibitem[{Sanders(2017)}]{sanders2017forecasting}
\bibinfo{author}{Sanders, N.R.}, \bibinfo{year}{2017}.
\newblock \bibinfo{title}{Forecasting: State-of-the-art in research and
  practice}, in: \bibinfo{booktitle}{The Routledge Companion to Production and
  Operations Management}. \bibinfo{publisher}{Routledge}, pp.
  \bibinfo{pages}{45--62}.
\bibitem[{Seabold and Perktold(2010)}]{seabold2010statsmodels}
\bibinfo{author}{Seabold, S.}, \bibinfo{author}{Perktold, J.},
  \bibinfo{year}{2010}.
\newblock \bibinfo{title}{Statsmodels: Econometric and statistical modeling
  with {p}ython}, in: \bibinfo{booktitle}{Proceedings of the 9th Python in
  Science Conference}, \bibinfo{organization}{Austin, TX}.
  p.~\bibinfo{pages}{61}.
\bibitem[{Shi et~al.(2020)Shi, Wang and Fang}]{shi2020artificial}
\bibinfo{author}{Shi, Z.R.}, \bibinfo{author}{Wang, C.}, \bibinfo{author}{Fang,
  F.}, \bibinfo{year}{2020}.
\newblock \bibinfo{title}{Artificial intelligence for social good: A survey}.
\newblock \bibinfo{journal}{arXiv preprint arXiv:2001.01818} .
\bibitem[{Singh(2016)}]{singh2016academia}
\bibinfo{author}{Singh, S.}, \bibinfo{year}{2016}.
\newblock \bibinfo{title}{Forecasting: Academia versus business}.
\newblock \bibinfo{journal}{Foresight: The International Journal of Applied
  Forecasting} , \bibinfo{pages}{46--47}.
\bibitem[{Van~Horn et~al.(2018)Van~Horn, Fierro, Kamdar, Gordon, Stewart,
  Bhattrai, Abe, Lei, O’Driscoll, Sinha et~al.}]{van2018democratizing}
\bibinfo{author}{Van~Horn, J.D.}, \bibinfo{author}{Fierro, L.},
  \bibinfo{author}{Kamdar, J.}, \bibinfo{author}{Gordon, J.},
  \bibinfo{author}{Stewart, C.}, \bibinfo{author}{Bhattrai, A.},
  \bibinfo{author}{Abe, S.}, \bibinfo{author}{Lei, X.},
  \bibinfo{author}{O’Driscoll, C.}, \bibinfo{author}{Sinha, A.}, et~al.,
  \bibinfo{year}{2018}.
\newblock \bibinfo{title}{Democratizing data science through data science
  training}, in: \bibinfo{booktitle}{Pacific {S}ymposium on {B}iocomputing
  2018: Proceedings of the Pacific Symposium}, \bibinfo{organization}{World
  Scientific}. pp. \bibinfo{pages}{292--303}.
\bibitem[{Vela(2018)}]{vela2018using}
\bibinfo{author}{Vela, K.}, \bibinfo{year}{2018}.
\newblock \bibinfo{title}{Using {S}lack to communicate with medical students}.
\newblock \bibinfo{journal}{Journal of the Medical Library Association: JMLA}
  \bibinfo{volume}{106}, \bibinfo{pages}{504}.
\bibitem[{Wang and Petropoulos(2016)}]{wang2016select}
\bibinfo{author}{Wang, X.}, \bibinfo{author}{Petropoulos, F.},
  \bibinfo{year}{2016}.
\newblock \bibinfo{title}{To select or to combine? the inventory performance of
  model and expert forecasts}.
\newblock \bibinfo{journal}{International Journal of Production Research}
  \bibinfo{volume}{54}, \bibinfo{pages}{5271--5282}.
\bibitem[{White et~al.(2011)White, Smith and Currie}]{white2011developing}
\bibinfo{author}{White, L.}, \bibinfo{author}{Smith, H.},
  \bibinfo{author}{Currie, C.}, \bibinfo{year}{2011}.
\newblock \bibinfo{title}{{OR} in developing countries: A review}.
\newblock \bibinfo{journal}{European Journal of Operational Research}
  \bibinfo{volume}{208}, \bibinfo{pages}{1--11}.
\bibitem[{Zahedi(1985)}]{zahedi1985ms}
\bibinfo{author}{Zahedi, F.}, \bibinfo{year}{1985}.
\newblock \bibinfo{title}{{MS}/{OR} education: Meeting the new demands on {MS}
  education}.
\newblock \bibinfo{journal}{Interfaces} \bibinfo{volume}{15},
  \bibinfo{pages}{85--94}.

\end{thebibliography}

\end{document}